\documentclass[fleqn,twoside]{article}
\usepackage{espcrc2}

\newcommand{\gtrsim}{\,\rlap{\lower3.7pt\hbox{$\mathchar\sim$}}
\raise1pt\hbox{$>$}\,}
\newcommand{\lesssim}{\,\rlap{\lower3.7pt\hbox{$\mathchar\sim$}}
\raise1pt\hbox{$<$}\,}

\usepackage{graphicx}
\usepackage{epsfig}

\usepackage[figuresright]{rotating}

\newcommand{\AmS}{{\protect\the\textfont2
  A\kern-.1667em\lower.5ex\hbox{M}\kern-.125emS}}

\title{Neutrino mass bounds from cosmology}

\author{Steen Hannestad \address{
Physics Department, University of Southern Denmark, Campusvej 55, \\ DK-5230 Odense M, Denmark, \\ and \\
Nordita, Blegdamsvej 17, DK-2100 Copenhagen, Denmark, \\ 
E-mail: hannestad@fysik.sdu.dk}}

\begin{document}

\begin{abstract}
Cosmology is at present one of the most powerful probes of neutrino properties. The advent of precision data from the cosmic microwave background and large scale structure has allowed for a very strong bound on the neutrino mass. Here, I review the status of cosmological bounds on neutrino properties with emphasis on mass bounds on light neutrinos.
\vspace{1pc}
\end{abstract}

\maketitle

Neutrinos are the among the most abundant particles in the
universe. This means that they have a profound impact on many different
aspects of cosmology, from the question of leptogenesis in the
very early universe, over big bang nucleosynthesis, to late time
structure formation. 

At late times ($T \lesssim T_{\rm EW}$) neutrinos mainly influence cosmology because of their energy density and, even later, their mass.

The absolute value of neutrino masses are very difficult to
measure experimentally. On the other hand, mass differences
between neutrino mass eigenstates, $(m_1,m_2,m_3)$, can be
measured in neutrino oscillation experiments.

The combination of all currently available data suggests two
important mass differences in the neutrino mass hierarchy. The
solar mass difference of $\delta m_{12}^2 \simeq 8 \times 10^{-5}$
eV$^2$ and the atmospheric mass difference $\delta m_{23}^2 \simeq
2.6 \times 10^{-3}$ eV$^2$
\cite{Maltoni:2004ei,Maltoni:2003da,Aliani:2003ns,deHolanda:2003nj}.

In the simplest case where neutrino masses are hierarchical these
results suggest that $m_1 \sim 0$, $m_2 \sim \delta m_{\rm
solar}$, and $m_3 \sim \delta m_{\rm atmospheric}$ \cite{giunti}. If the
hierarchy is inverted
one instead finds $m_3 \sim 0$, $m_2 \sim \delta m_{\rm
atmospheric}$, and $m_1 \sim \delta m_{\rm atmospheric}$. However,
it is also possible that neutrino masses are degenerate, $m_1 \sim m_2 \sim m_3 \gg \delta
m_{\rm atmospheric}$, in which case oscillation experiments are
not useful for determining the absolute mass scale \cite{giunti}.

Experiments which rely on kinematical effects of the neutrino mass
offer the strongest probe of this overall mass scale. Tritium
decay measurements have been able to put an upper limit on the
electron neutrino mass of 2.3 eV (95\% conf.) \cite{kraus} (see also contribution by G.~Drexlin to the present volume).
However, cosmology at present yields a much stronger limit which
is also based on the kinematics of neutrino mass.

Very interestingly there is also a claim of direct detection of
neutrinoless double beta decay in the Heidelberg-Moscow experiment
\cite{Klapdor},
corresponding to an effective neutrino mass in the $0.1-0.9$ eV
range. If this result is confirmed then it shows that neutrino
masses are almost degenerate and well within reach of cosmological
detection in the near future.

Neutrinos are not the only possibility for stable eV-mass particles in the universe. There are numerous other candidates, such as axions and majorons, which might be present. As will be discussed later, the same cosmological mass bounds can be applied to any generic light particle which was once in thermal equilibrium.

Here I focus mainly on the issue of cosmological mass bounds. Much more detailed reviews of neutrino cosmology can for instance be found in \cite{dolg,sth}.

\section{Neutrinos in structure formation}

The temperature of standard model neutrinos is given by $T_\nu = (4/11)^{1/3} T_\gamma$, and the number density of a given flavour is therefore $n_\nu = 7/8 \times (4/11) \times n_\gamma$. From this the present contribution to the matter density from massive neutrinos is
\begin{equation}
\Omega_\nu h^2 = N_\nu \frac{m_\nu}{92.5 \,\, {\rm eV}},
\end{equation}
where $N_\nu$ is the number of neutrino species.
Thus, even a sub-eV neutrino mass gives a significant neutrino contribution to the energy density and therefore has an effect on structure formation.

Perturbations in the neutrino distribution can be followed by solving the Boltzmann equation \cite{mb} through the epoch of structure formation. There are publicly available codes, such as CMBFAST \cite{CMBFAST}, which can do this.

\begin{figure}
\begin{center}
\hspace*{-0.4cm}
\epsfig{file=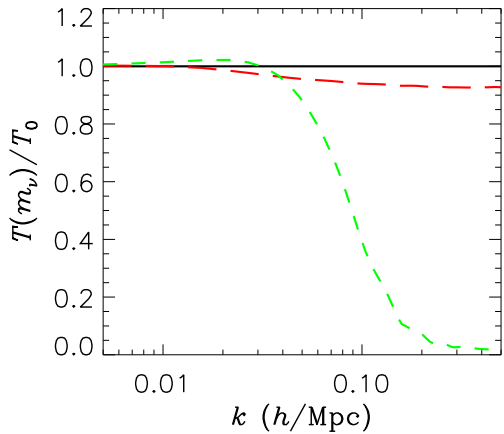,width=0.5\textwidth}
\end{center}
\bigskip
\caption{\label{fig:nutrans} The transfer function $T(k,t=t_0)$
for various different neutrino masses. The solid (black) line is
for $m_\nu=0$, the long-dashed for $m_\nu = 0.3$ eV, and the
dashed for $m_\nu=1$ eV.}
\end{figure}

The main difference between neutrinos and cold dark matter is that neutrinos are light and only become non-relativistic around the epoch of recombination, significantly later than matter-radiation equality.
While they are relativistic, neutrinos free-stream a distance roughly given by $\lambda = 2\pi/k = c \tau$, where $\tau$ is conformal time. The total free-streaming length is therefore roughly given by $\lambda_f \sim c \tau(T = m_\nu)$. At scales smaller than this, the solution to the Boltzmann equation is exponentially damped. For scales larger than the free-streaming length neutrino perturbations are unaffected. 
It is therefore possible to divide the solution into two distinct regimes:
1) $k > \tau(T=m)$: Neutrino perturbations are exponentially damped
2) $k < \tau(T=m)$: Neutrino perturbations follow the CDM
perturbations.
Calculating the free streaming wavenumber in a flat CDM cosmology
leads to the simple numerical relation (applicable only for
$T_{\rm eq} \gg m \gg T_0$)

\begin{eqnarray}
\lambda_{\rm FS} & \sim & \frac{20~{\rm Mpc}}{\Omega_x h^2} \left(\frac{T_x}{T_\nu}\right)^4 \nonumber \\
&& \,\, \times  \left[1+\log \left(3.9
\frac{\Omega_x h^2}{\Omega_m h^2} \left(\frac{T_\nu}{T_x}\right)^2
\right)\right]\,. \label{eq:freestream}
\end{eqnarray}

In Fig.~\ref{fig:nutrans} transfer functions for
various different neutrino masses in a flat $\Lambda$CDM universe
$(\Omega_m+\Omega_\nu+\Omega_\Lambda=1)$ are plotted. 
The parameters used were
$\Omega_b = 0.04$, $\Omega_{\rm CDM} = 0.26 - \Omega_\nu$,
$\Omega_\Lambda = 0.7$, $h = 0.7$, and $n=1$.

When measuring fluctuations it is customary to use the power
spectrum, $P(k,\tau)$, defined as
\begin{equation}
P(k,\tau) = |\delta_k(\tau)|^2.
\end{equation}
The power spectrum can be decomposed into a primordial part,
$P_0(k)$, and a transfer function $T(k,\tau)$,
\begin{equation}
P(k,\tau) = P_0(k) T(k,\tau).
\end{equation}
The transfer function at a particular time is found by solving the
Boltzmann equation for $\delta(\tau)$.

At scales much smaller than the free-streaming scale the present
matter power spectrum is suppressed roughly by the factor
\cite{Hu:1997mj}
\begin{equation}
\frac{\Delta P(k)}{P(k)} = \frac{\Delta
T(k,\tau=\tau_0)}{T(k,\tau=\tau_0)}\simeq -8
\frac{\Omega_\nu}{\Omega_m},
\end{equation}
as long as $\Omega_\nu \ll \Omega_m$. The numerical factor 8 is
derived from a numerical solution of the Boltzmann equation, but
the general structure of the equation is simple to understand. At
scales smaller than the free-streaming scale the neutrino
perturbations are washed out completely, leaving only
perturbations in the non-relativistic matter (CDM and baryons).
Therefore the {\it relative} suppression of power is proportional
to the ratio of neutrino energy density to the overall matter
density. Clearly the above relation only applies when $\Omega_\nu
\ll \Omega_m$, when $\Omega_\nu$ becomes dominant the spectrum
suppression becomes exponential as in the pure hot dark matter
model. This effect is shown for different neutrino masses in
Fig.~\ref{fig:nutrans}.

The effect of massive neutrinos on structure formation only
applies to the scales below the free-streaming length. For
neutrinos with masses of several eV the free-streaming scale is
smaller than the scales which can be probed using present CMB data
and therefore the power spectrum suppression can be seen only in
large scale structure data. On the other hand, neutrinos of sub-eV
mass behave almost like a relativistic neutrino species for CMB
considerations. The main effect of a small neutrino mass on the
CMB is that it leads to an enhanced early ISW effect. The reason
is that the ratio of radiation to matter at recombination becomes
larger because a sub-eV neutrino is still relativistic or
semi-relativistic at recombination. With the WMAP data alone it is
very difficult to constrain the neutrino mass, and to achieve a
constraint which is competitive with current experimental bounds
it is necessary to include LSS data from 2dF or SDSS. When this is
done the bound becomes very strong, somewhere in the range of 1 eV
for the sum of neutrino masses, depending on assumptions about
priors. 
This bound can be strengthened even further by including data from the Lyman-$\alpha$ forest and assumptions about bias. In this case the bound on the sum of neutrino masses becomes as low as 0.4-0.6 eV.

\begin{figure}[htb]
\begin{center}
\hspace*{-0.4cm}
\epsfig{file=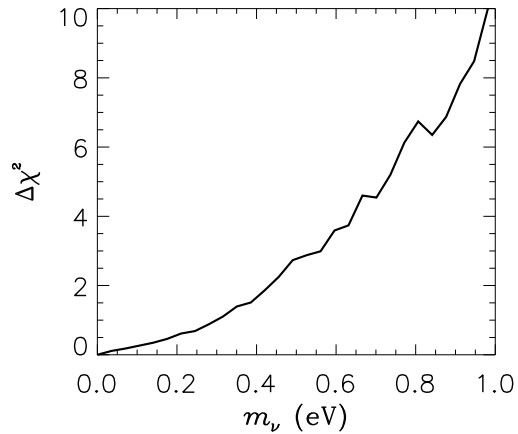,width=0.5\textwidth}
\end{center}

\caption{$\Delta \chi^2$ as a function of the the sum of neutrino masses $m_\nu$.}
\label{fig:like}
\end{figure}

In Fig.~\ref{fig:like} $\Delta \chi^2$ is shown for an analysis which includes the WMAP CMB data \cite{WMAP}, the SDSS galaxy survey data \cite{Tegmark:2003uf,Tegmark:2003ud}, the Riess {\it et al.} SNI-a "gold" sample \cite{Riess:2004}, and the Lyman-$\alpha$ forest data from Croft {\it et al.} \cite{lya}. For the Lyman-$\alpha$ forest data, the error bars on the last three data points have been increased in the same fashion as was done by the WMAP collaboration \cite{WMAP}, in order to make them compatible with the analysis of Gnedin and Hamilton \cite{gneha}. In addition to the neutrino mass, which I take to be distributed in three degenerate species, I take the
minimum standard model with 6
parameters: $\Omega_m$, the matter density, $\Omega_b$, the baryon density, $H_0$, the Hubble
parameter, and $\tau$, the optical depth to reionization. The normalization of both
CMB, LSS, and Ly-$\alpha$ spectra are taken to be free and unrelated parameters.
The priors used are given in Table~\ref{tab:priors}.

\begin{table}[htb]
\caption{Priors on cosmological parameters used in
the likelihood analysis.}
\begin{tabular}{@{}lll@{}}
\hline
Parameter &Prior&Distribution\cr
\hline
$\Omega=\Omega_m+\Omega_X$&1&Fixed\\
$h$ & $0.72 \pm 0.08$&Gaussian \cite{freedman}\\
$\Omega_b h^2$ & 0.014--0.040&Top hat\\
$n_s$ & 0.6--1.4& Top hat\\
$\tau$ & 0--1 &Top hat\\
$Q$ & --- &Free\\
$b$ & --- &Free\\
\hline
\end{tabular}\label{tab:priors} 
\end{table}

In this particular analysis, the 95\% C.L. upper bound on the sum of neutrino masses is 0.55 eV.

In Table \ref{table:massnu} the present upper bound on the
neutrino mass from various analyses is quoted, as well as the
assumptions going into the derivation.
As can be gauged from this table, a fairly robust bound on the sum
of neutrino masses is at present somewhere around 0.5-1 eV,
depending on the specific priors and data sets used.

\begin{table*}
\caption{Various recent limits on the neutrino mass from cosmology
and the data sets used in deriving them. 1: WMAP data, 2: Other
CMB data, 3: 2dF data, 4: Constraint on $\sigma_8$ (different in
$4^a$, $4^b$, and $4^c$), 5: SDSS data, 6: Constraint on $H_0$, 7: Constraint from Lyman-$\alpha$ forest.}
\begin{tabular}{@{}lll@{}}
\hline
Ref. & Bound on $\sum m_\nu$ & Data used \\
\hline
Spergel et al. (WMAP) \cite{WMAP}  &   0.69 eV     &   1,2,3,$4^a$,6, 7 \\
Hannestad \cite{Hannestad:2003xv} &   1.01 eV     &   1,2,3,6 \\
Allen, Smith and Bridle \cite{Allen:2003pt} & $0.56^{+0.3}_{-0.26}$ eV & 1,2,3,$4^b$,6 \\
Tegmark et al. (SDSS) \cite{Tegmark:2003ud} & 1.8 eV & 1,5 \\
Barger et al. \cite{Barger:2003vs} & 0.75 eV & 1,2,3,5,6 \\
Crotty, Lesgourgues and Pastor \cite{Crotty:2004gm} & 1.0 (0.6) eV & 1,2,3,5 (6) \\
Seljak et al. \cite{sel04} & 0.42 eV & 1,2,$4^c$,5,6,7 \\
Fogli et al. \cite{fogli04} & 0.5 eV & 1,2,3,4$^c$,5,6,7 \\
The present work \cite{seesaw} & 0.65 eV & 1,5,6,7 \\

\hline
\end{tabular}
\label{table:massnu}
\end{table*}

\section{General thermal relics}

While light neutrinos are the canonical light, thermally produced 
particles, there are other species which can have the same properties. At $T \sim M_{\rm Pl}$ gravitons are presumably in thermal equilibrium. However, since inflation occurs at a lower temperature, there should be no thermal graviton background present.

Other examples include majorons, axions, etc. Any such species which was once in equilibrium is characterized by only two quantities, its mass, $m_X$, and the temperature at which it decoupled from thermal equilibrium, $T_{\rm D}$. At this temperature the number of degrees of freedom was $g_{*}$. The relevant masses are $\ll$ MeV so that the particles
decouple when they are relativistic, i.e.\ at decoupling they are
characterized by a Fermi-Dirac or Bose-Einstein distribution of
temperature $T_{\rm D}$. 

In the present-day universe, the new particles will be
non-relativistic and contribute a matter fraction
\begin{equation}\label{eq:omegax}
\Omega_X h^2=\frac{m_X g_X}{183~{\rm eV}}\,
\frac{g_{*\nu}}{g_*}\times \cases{
1 & {\rm fermions} \cr \frac{4}{3} & {\rm bosons} }
\end{equation}
where $g_X$ is the number of the particle's internal degrees of
freedom while $g_{*\nu}$ is the effective number of thermal degrees of
freedom when ordinary neutrinos freeze out with $g_{*\nu}=10.75$ in
the absence of new particles.  

In the late epochs that are important for structure formation, the
momentum distribution of the new particles is characterized by a
thermal distribution with temperature $T_X$ that is given by
\begin{equation}
\frac{T_X}{T_\nu}=\left(\frac{g_{*\nu}}{g_*}\right)^{1/3}
\,.
\end{equation}

In Ref.~\cite{HR04} CMBFAST was modified to incorporate such thermal relics, both fermionic and bosonic.

\begin{figure}[htb]
\begin{center}
\hspace*{-0.4cm}
\epsfig{file=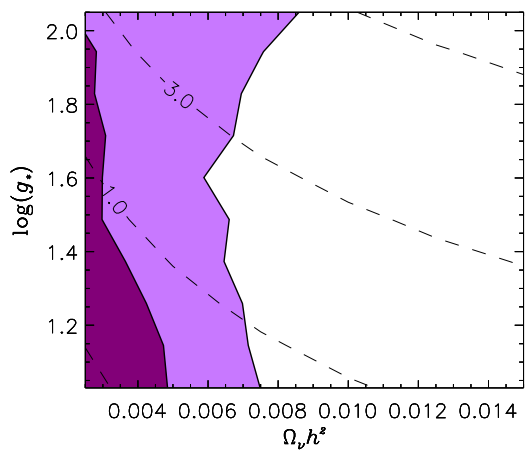,width=0.5\textwidth}
\end{center}

\caption{The full lines show 
68\% and 95\% confidence regions in the $(\Omega_\nu h^2,g_*)$
plane for the case where the thermal relic is a fermion with $g=2$.}
\label{fig:relic}
\end{figure}

Here I present an updated analysis for the fermionic case with $g_X=2$ (i.e.\ a single Majorana fermion), using the same data sets as described above. Fig.~\ref{fig:relic} shows the 68\% and 95\% likelihood contours for $\Omega_X h^2$ and $g_*$. Overlayed are isocontours for particle masses (in eV).

As an example, a species freezing out at the electroweak transition temperature has $g_* = 106.75$, and from the figure it can be seen that the upper bound on the mass of such a particle is around 5 eV. For a corresponding scalar the mass bound will be slightly higher. For a species decoupling after the QCD phase transition where $g_* \lesssim 20$ the mass bound is roughly $m \lesssim 1$ eV.

\section{Conclusion}

I have reviewed the present status of cosmological mass bounds on neutrinos and other light, thermally produced particles. Already now these bounds are about an order of magnitude stronger than current laboratory limits on the neutrino mass, albeit more model dependent.
In the coming years a wealth of new cosmological data will become available from such experiments as the Planck Surveyor CMB satellite. This is likely to allow for a measurement of the (sum of) neutrino masses in the 0.1 eV regime \cite{Hannestad:2002cn,Lesgourgues:2004ps,Kaplinghat:2003bh,Abazajian:2002ck}
. Together with direct detection experiments like KATRIN \cite{katrin} and future neutrinoless double beta decay experiments, cosmology will provide the answer to whether neutrino masses are hierarchical.

\end{document}